\documentclass[12pt]{article}

\usepackage{newtxtext,newtxmath}

\usepackage{graphicx}
\usepackage[letterpaper,margin=1in]{geometry}

\renewenvironment{abstract}
	{\quotation}
	{\endquotation}

\date{}

\makeatletter
\renewcommand{\fnum@figure}{\textbf{Figure \thefigure}}
\renewcommand{\fnum@table}{\textbf{Table \thetable}}
\makeatother

\usepackage{scicite}
\usepackage{url}

\def\scititle{
	Isostructural electronic transition in MoS$_2$ probed by solid-state high harmonic generation spectroscopy
}
\title{\bfseries \boldmath \scititle}

\author{
	Bailey~R.~Nebgen$^{1,2}$,
	Victor~Chang~Lee$^{3}$,
	Jacob~A.~Spies$^{1,2}$,
        Randy~M.~Sterbentz$^{4}$\and
        Craig~P.~Schwartz$^{5}$,
        Dean~Smith$^{5,6}$,
        Diana~Y.~Qiu$^{3}$,
        Michael~W.~Zuerch$^{1,2\ast}$\and
	\small$^{1}$Department of Chemistry, University of California Berkeley, Berkeley, CA 94720, USA.\and
	\small$^{2}$Materials Sciences Division, Lawrence Berkeley National Laboratory, Berkeley, CA 94720, USA.\and
	\small$^{3}$Department of Materials Science, Yale University, New Haven, CT 06520, USA.\and
	\small$^{4}$Department of Physics and Astronomy, University of Nevada Las Vegas, Las Vegas, NV 89154, USA.\and
	\small$^{5}$Nevada Extreme Conditions Laboratory, University of Nevada Las Vegas, Las Vegas, NV 89154, USA.\and
	\small$^{6}$High Pressure Collaborative Access Team (HPCAT), X-Ray Science Division, Argonne National Laboratory, \\ \small Lemont, IL 60439, USA.\and
	\small$^\ast$Corresponding author. Email: mwz@berkeley.edu\and
}

\begin{document} 

\maketitle

\begin{abstract} \bfseries \boldmath

Studying materials under extreme pressure in diamond anvil cells (DACs) is key to discovering new states of matter, yet no method currently allows the direct measurement of the electronic structure in this environment. Solid-state high harmonic generation (sHHG) offers a new all-optical window into the electronic structure of materials. We demonstrate sHHG spectroscopy inside a DAC by probing $2H$-MoS$_2$, up to 30 GPa, revealing a pressure-induced crossover of the lowest direct bandgap from the \textbf{K}-point to the $\Gamma$-point. This transition manifests as a sharp minimum in harmonic intensity and a 30° rotation of the sHHG polarization anisotropy, despite the absence of a structural phase change. First-principles simulations attribute these features to interference between competing excitation pathways at distinct points in the Brillouin zone. Our results establish sHHG as a sensitive probe of electronic transitions at high pressure, enabling access to quantum phenomena that evade detection by conventional techniques.

\end{abstract}

\noindent
Extreme pressure affords access to states, structures, and chemical compounds which are inaccessible at ambient conditions, making it a powerful platform for the discovery of new functional materials.~\cite{McMillan2002}
In condensed matter, pressure offers a method for tuning electronic properties of materials by directly modulating interatomic spacing. The effects of pressure can subsequently be mimicked by chemical doping or strain engineering~\cite{Castellanos-Gomez2013}, positioning pressure-dependent studies of materials as a strategic route for designing materials for emerging technologies.
Diamond anvil high-pressure cells (DAC) are a commonly used platform for both generating extreme pressures owing to the hardness of diamond, and for spectroscopic characterization of materials under pressure owing to the simple chemical makeup and crystal structure of diamond which results in a large bandgap in the UV and a broadband optical transparency with few absorption features.
The overlap between DAC high pressure experiments and ultrafast spectroscopy is a burgeoning field, with recent results from pump-probe spectroscopy giving insights spanning thermal properties of simple metals~\cite{Hsieh2024} to electronic properties of high-pressure hydrides and nickelates~\cite{Wu2024, Meng2024}.

\begin{figure}
	\centering
	\includegraphics[width=1\textwidth]{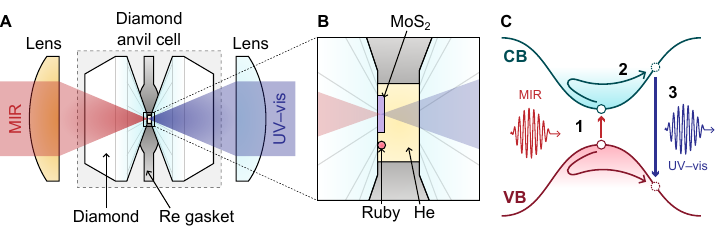}
	\caption{
    \textbf{Solid state high harmonic generation spectroscopy at high pressure in a DAC.}
    (\textbf{A)} Mid-infrared (MIR; red) light is incident on the MoS$_2$ sample through one diamond window in a diamond anvil cell (DAC), high harmonic spectra (UV--vis; blue) are collected in transmission geometry through the other diamond window.
    (\textbf{B)} DAC sample chamber showing MoS$_{2}$ flake deposited on culet surface of diamond, alongside a ruby sphere used as a pressure marker. The sample chamber is sealed in an atmosphere of He at 3~kbar to provide hydrostatic compression.\cite{Klotz2009}
    (\textbf{C)} The three-step mechanism of sHHG in a schematic of a semiconductor band structure. First (1), the MIR pulse tunnel-excites an electron to the conduction band (CB).
    Then (2), the electron and hole are accelerated by the MIR pulse, generating a nonlinear current in their respective bands before (3) recombination.
    Both steps 2 and 3 lead to UV-vis sHHG emissions.\cite{Ghimire2018}
    }
	\label{fig:DACschematic}
\end{figure}

However, currently a major limitation for extreme pressure experiments is the difficulty of performing momentum-resolved measurements of the electronic structure, as they are routinely performed in ultra-high vacuum, e.g.,  using angle-resolved photoemission spectroscopy (ARPES)~\cite{Zhang2022}. More generally, powerful methods that rely on photoemission of electrons become inapplicable, as the emitted electrons cannot traverse the diamond anvils required to sustain high pressures.
In contrast, solid-state high harmonic generation (sHHG) spectroscopy is an emerging all-optical technique that can probe electronic structure~\cite{Heide2024} including properties such as symmetry~\cite{SingYou2017, Yue2022signatures, He2022}, dynamics~\cite{Langer2016, Silva2018, Bionta2021}, topology~\cite{Baykusheva2021, Heide2022topological}, and band dispersion~\cite{Vampa2015, Parks2025}.
Compared to traditional perturbative (i.e., multi-photon) nonlinear spectroscopies, sHHG spectroscopy utilizes a strong-field mid-infrared laser pulse to generate high-order harmonics of the fundamental driving field frequency in the non-perturbative regime~\cite{Ghimire2011, Zong2023}.
The emitted high-order harmonics result from nonlinear currents driven within bands by the driving field as well as interband recombination (Fig.~\ref{fig:DACschematic}C)~\cite{Ghimire2018, SingYou2017}.
Harmonic generation efficiency and thus observed spectra are highly sensitive to elements of the electronic structure such as bandgap, band dispersion, and carrier population. Accordingly, many aspects of the momentum-resolved band structure can potentially be extracted from sHHG~\cite{Vampa2015, Parks2025}. 
Critical information is often extracted using sHHG anisotropy, where harmonic emissions are measured as dependent on the angle between the driving field polarization and the crystal structure being studied~\cite{Liu2017, SingYou2017, Yue2022signatures}.
Measuring the polarization anisotropy of sHHG emission allows access to a detailed picture of electronic and structural phenomena because the angular dependence is constrained by crystal symmetry and is highly sensitive to the landscape of the band structure~\cite{SingYou2017, Yue2022signatures}.
By capitalizing on the broadband transparency of diamond in the mid-infrared as well the visible to ultraviolet range, sHHG is thus a promising technique providing an all-optical probe of momentum-resolved electronic structure of materials under extreme pressures in a DAC (Fig.~\ref{fig:DACschematic}A and B). Despite its promise, sHHG under extreme pressure has only been theoretically explored in a single study, which analyzed its potential to probe exotic quantum states in sulfur hydrides~\cite{Hu2024}.

Here, we compressed bulk MoS$_2$ in a diamond anvil cell to pressures up to 30~GPa and demonstrated sHHG using a mid-infrared driving field at a photon energy of 0.38~eV to generate harmonics up to the 13$^{\text{th}}$ order. MoS$_2$ was the first two-dimensional material experimentally shown to exhibit sHHG~\cite{Liu2017}, and has since been extensively studied under ambient conditions, including detailed investigations of the polarization-dependent characteristics and anisotropic behavior of its sHHG response~\cite{Yoshikawa2019, Lou2020, Yue2022signatures, Heide2022coherence, ChangLee2024}. Combined with its relatively simple quasi-two-dimensional structure, this makes MoS$_2$ an ideal model system for evaluating the capabilities of sHHG as a probe of electronic structure under extreme pressure. To the best of our knowledge, this work demonstrates the first measurement of sHHG emissions from a material inside a DAC, opening a pathway to obtain experimental insights into electronic structure at extreme pressure.
We observed marked and non-monotonic changes in both the intensity and anisotropy of the harmonic emission as a function of pressure, despite the absence of known structural phase transitions in this pressure range. The observed intensity and anisotropy variations reflect a pressure-induced reorganization of the electronic structure, consistent with an isostructural electronic transition near 15~GPa. Through comparison with first-principles band structure calculations and simulations of the sHHG response, we show that the evolution of the emission characteristics correlates with a crossover in the bandgap minimum from the \textbf{K}-point to the $\mathbf{\Gamma}$-point. The ability to track this crossover through purely optical means establishes sHHG as a powerful and momentum-sensitive probe of electronic structure under extreme pressure, offering access to transitions that are invisible to conventional structural probes and broadly enabling high-pressure studies of many materials including quantum materials.

\subsection*{Extreme pressure sHHG measurements}

\begin{figure}
    \centering
    \includegraphics[width=1\textwidth]{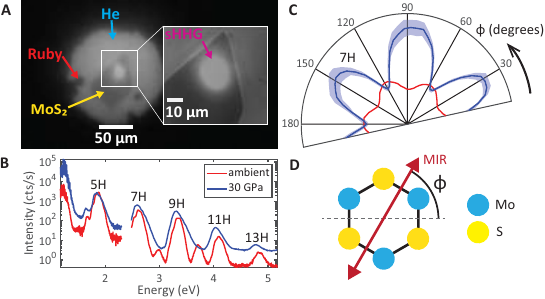}
    \caption{
    \textbf{sHHG emissions from MoS$_2$ inside a DAC.}
    (\textbf{A}) Microscope image of the exfoliated bulk MoS$_2$ flake inside a DAC.
    The MIR driver was incident on the flake and harmonic emissions are visible as a roughly circular light area in the image.
    The inset was taken with a 50x objective and shows how the sample was positioned during the high pressure sHHG measurements.
    (\textbf{B}) Representative sHHG spectra of MoS$_2$ inside the DAC at ambient pressure (red) and 30~GPa (blue).
    Spectra are averaged over all polarization angles of the MIR driver.
    (\textbf{C}) Anisotropy of the 7$^{\text{th}}$ harmonic from MoS$_2$ inside the DAC at ambient pressure (red) and 30~GPa (blue).
    Intensity at each angle represents Gaussian amplitude after fitting the peak of each harmonic.
    The shaded region represents the standard deviation of the fitted Gaussian amplitudes of several measurements.
    (\textbf{D}) A schematic of the anisotropy measurement relative to the 2D-projection of the structure of MoS$_2$ where the angle gives the orientation of the polarization of the MIR driver relative to the crystal.
    Anisotropy measurements presented later use the same definition of the angle $\phi$.
    }
	\label{fig:DACexample}
\end{figure}

Harmonic emissions from MoS$_2$ inside a DAC are shown in Fig.~\ref{fig:DACexample} and the experimental apparatus is shown in detail in Fig.~\ref{fig:setup}B and described in the Methods.
Briefly, a linear-polarized driving MIR pulse of $\sim$100~fs centered at 0.38~eV was attenuated using a pair of polarizers and focused onto the MoS$_2$ flake with a focal spot diameter of 30~$\mu$m.
sHHG spectra in this study were generated with a MIR intensity of 0.71$\pm$0.08~TW/cm$^2$ and spectra were measured using a UV-vis spectrometer.
Pressure was measured by inducing fluorescence from rubies included in the sample chamber of the DAC and comparing to rubies at ambient pressure (see Methods)~\cite{Dewaele2008}.
Harmonic emissions were first confirmed to originate from MoS$_2$ inside the DAC using a built-in imaging system as described in the Methods (Fig.~\ref{fig:setup}A). With the imaging camera blind to the MIR driver, the sHHG emissions in the UV-visible range were easily observable as a white emission feature (Fig.~\ref{fig:DACexample}A), allowing the MIR beam to be centered on the sample.
The harmonic spectrum from MoS$_2$ inside the DAC (Fig.~\ref{fig:DACexample}B) was then compared to background signal from the surrounding diamond, which was significantly weaker due to its large bandgap (Fig.~\ref{fig:backgroundSignal}).
As described further in the Methods, most harmonics from MoS$_2$ are free of significant background diamond signal except for the 5$^{\text{th}}$ harmonic which was excluded from the analysis.
sHHG from MoS$_2$ was measured at pressures up to 30~GPa, as well as with the DAC opened to atmosphere. The spectrum at 30~GPa demonstrates full suppression of even harmonics due to bulk centrosymmetry, helping to confirm sample uniformity at extreme pressure (Fig.~\ref{fig:DACexample}B blue line).
For a direct comparison, the sHHG spectrum at ambient pressure is compared to 30~GPa in Fig.~\ref{fig:DACexample}B, showing odd harmonics out to the 13$^{\text{th}}$ in both cases and an overall increase in signal at higher pressure.

The anisotropy of the harmonics was measured inside a DAC by rotating the polarization of the MIR driver relative to the MoS$_2$ sample. Due to potential effects of diamond on the MIR driver polarization, special care was taken to avoid misinterpretation of the sHHG anisotropy (see Methods). At ambient pressure, the harmonics clearly showed the expected six-fold symmetry that has been previously observed from MoS$_2$ (Fig.~\ref{fig:DACexample}C, red line)~\cite{Liu2017, Yue2022signatures}.
However, as shown by the blue line in Fig.~\ref{fig:DACexample}C, the anisotropy was significantly altered at extreme pressure. By 30~GPa, the 7$^{\text{th}}$ harmonic no longer exhibited perfectly six-fold symmetry, and the observed anisotropy pattern had rotated by 30$^\circ$ relative to the measurement at ambient pressure.
While one might expect that this dramatic change in sHHG anisotropy indicates a significant structural phase transition, previous studies have shown that MoS$_2$ maintains a $2H$ structure throughout the studied pressure range, but may slide into a different interlayer stacking configuration~\cite{Aksoy2006, Hromadova2013, Bandaru2014, Fan2015}.
Specifically, it has been shown and reproduced that at ambient conditions MoS$_2$ is in a $2H_c$ stacking configuration where Mo atoms in consecutive layers are horizontally offset and as pressure increases, all of the lattice parameters decrease as noted in Table~S1~\cite{Aksoy2006, Hromadova2013, Bandaru2014}. Around 26~GPa, MoS$_2$ begins to have some phase contribution of $2H_a$ stacking configuration where the Mo atoms are vertically aligned along the c-axis~\cite{Aksoy2006, Hromadova2013, Bandaru2014}.
In order to determine whether this interlayer sliding transition or another effect is the cause of the drastic change to the sHHG anisotropy, a detailed pressure-dependent anisotropy study was performed.

\begin{figure}
    \centering
    \includegraphics[width=0.7\textwidth]{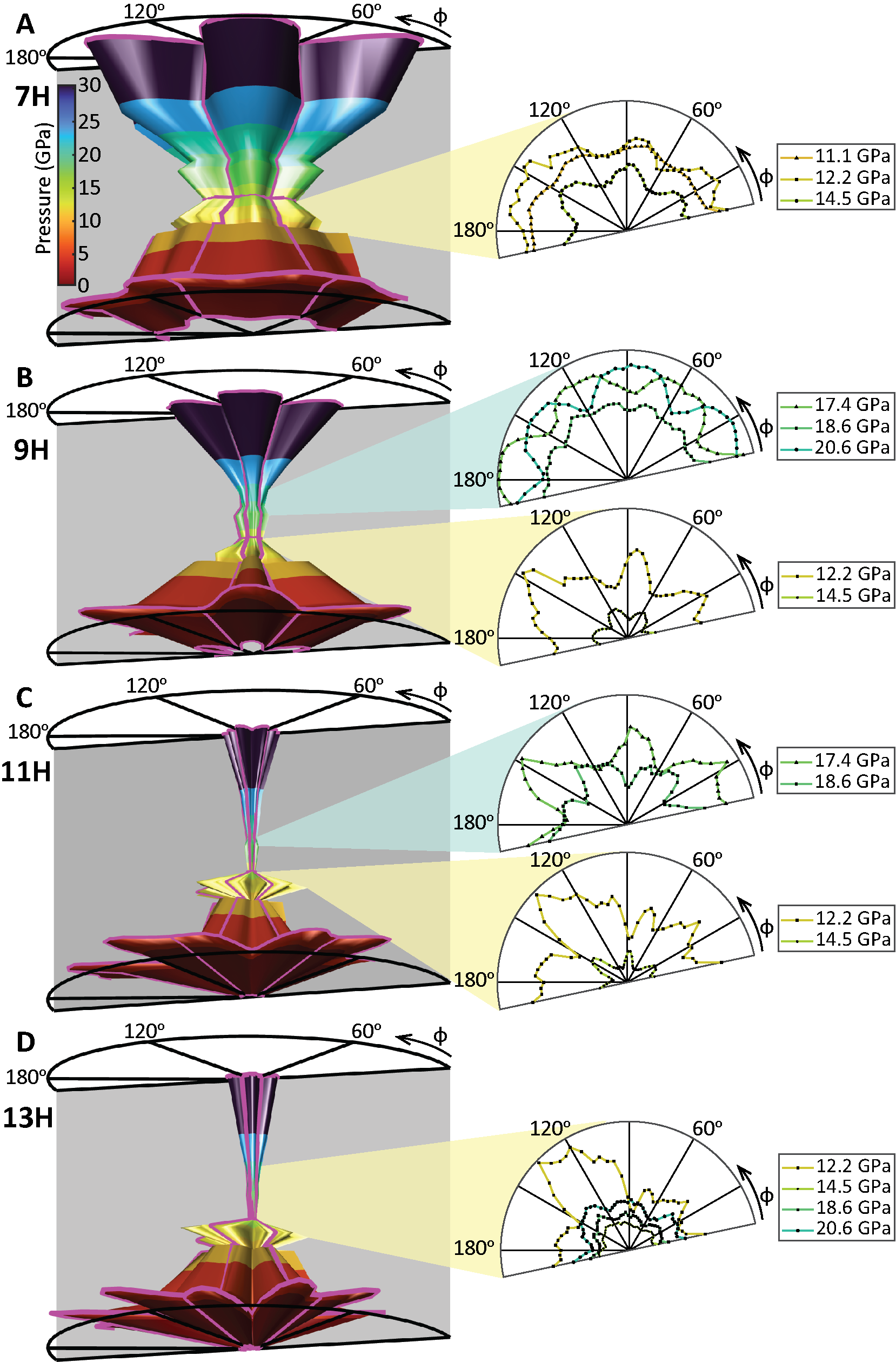}
    \caption{
    \textbf{Pressure-dependent sHHG anisotropy of MoS$_2$.}
	3D contours of the pressure-dependent anisotropic sHHG emissions from MoS$_2$ in cylindrical coordinates with the vertical axis and color corresponding to pressure and the angular axis corresponding to MIR driver polarization relative to the crystal axes of MoS$_2$ as defined previously.
    Intensity of the (\textbf{A}) 7$^{\text{th}}$ harmonic, (\textbf{B}) 9$^{\text{th}}$ harmonic, (\textbf{C}) 11$^{\text{th}}$ harmonic, and (\textbf{D}) 13$^{\text{th}}$ harmonic as dependent on polarization angle and pressure.
    The insets to the right highlight certain pressure ranges where the anisotropy changes significantly.
    }
    \label{fig:pressureAnisotropy}
\end{figure}

Anisotropies of the 7$^{\text{th}}$, 9$^{\text{th}}$, 11$^{\text{th}}$ and 13$^{\text{th}}$ harmonics demonstrate changes across the incremental pressure range of 0-30~GPa which report on changes to the electronic structure (Fig.~\ref{fig:pressureAnisotropy}).
While each harmonic behaved slightly differently, there are several common features.
First, each harmonic started at ambient conditions with six-fold symmetric anisotropy that was later modified to varying extents by the applied pressure.
Most strikingly, the intensities of all harmonics exhibited a minimum at 14.5$\pm$0.1~GPa, outside the pressure range at which the interlayer sliding transition has been shown to occur~\cite{Aksoy2006, Hromadova2013, Bandaru2014}.
In addition, while some minor deviation from perfect six-fold symmetry may occur at lower pressures, all harmonics underwent a significant change in symmetry around the same pressure at which the minimum in emissions occurs.
This behavior is highlighted in the insets of Fig.~\ref{fig:pressureAnisotropy}, which show detailed views of the pressure ranges where the angular dependence of the harmonics change most significantly. The 7$^{\text{th}}$, 9$^{\text{th}}$, and 11$^{\text{th}}$ harmonic anisotropies shifted their angular maxima by 30$^\circ$, while the 13$^{\text{th}}$ harmonic became mostly isotropic by 14.5$\pm$0.1~GPa. This commonality in behavior at 14.5~GPa suggests a significant change in the electronic structure affecting all electron trajectories in the sHHG mechanism despite the $2H_c$ phase remaining constant at this pressure. In addition, by about 20.6~GPa, the 9$^{\text{th}}$ and 11$^{\text{th}}$ harmonic anisotropies underwent a second angular shift back to an orientation similar to ambient conditions and the 13$^{\text{th}}$ harmonic rotated 30$^\circ$ from its ambient orientation. These continued changes in anisotropic symmetry up to 20.6~GPa suggest suggest continued pressure-induced changes in electronic structure. Above 20.6~GPa, sHHG emission increased monotonically while maintaining similar anisotropy, which suggests bandgap closing without significant warping of the electronic structure landscape.

The observed angular shifts in anisotropy between $\sim$12-20~GPa may seem surprising at first given the lack of a structural phase transition in the pressure range.
However, similar 30$^\circ$ angular shifts in the anisotropy of odd harmonics from MoS$_2$ caused by electronic structure effects have been extensively investigated previously~\cite{Yue2022signatures}.
Yue \textit{et al.} showed that within a single sHHG anisotropy measurement of monolayer MoS$_2$, odd parallel harmonics were angularly shifted by 30$^\circ$ relative to each other dependent on the energy of the harmonic~\cite{Yue2022signatures}.
This rotation was found to directly report on harmonics being emitted from different critical points within the band structure. Specifically, the mechanism for the lower energy harmonics was dominated by the first CB, while higher energy harmonics above the gap to the second CB of 3.5~eV were altered by the interference of emission channels in the first and second CB~\cite{Yue2022signatures}.
A similar effect is reproduced in the sHHG anisotropy of bulk MoS$_2$ in this experiment, shown by the 30$^\circ$ offset between the 7$^{\text{th}}$ and 9$^{\text{th}}$ harmonics at lower pressures (Fig.~\ref{fig:pressureAnisotropy}A and B). Given the previous in-depth analysis of the angular shift of odd harmonics from MoS$_2$ along with previous studies showing that MoS$_2$ should remain in the $2H_c$ structure in the pressure range where the sHHG anisotropy rotates, it suggests that the observed angular shifts in sHHG anisotropy with pressure likely represent changes to the electronic structure of MoS$_2$ that allow the sHHG carrier trajectories to access different critical points in the band structure.

\subsection*{Theoretical investigation}

Given the complex nature of the sHHG anisotropy in MoS$_2$ under extreme pressure and the sensitivity of the technique to multiple electronic factors, it is essential to theoretically evaluate the pressure-dependent band structure to interpret the observed emission features. This was achieved by first computing the electronic band structure as a function of pressure and then simulating the corresponding sHHG response to assess consistency with the experimental observations.
Here, the ambient band structure of MoS$_2$ was obtained by relaxing the atomic structure of $2H_c$-MoS$_2$ at 0~GPa using density functional theory (DFT)~\cite{Kohn1968} with the Perdew-Burke-Ernzerhof (PBE)~\cite{Perdew1997} exchange-correlation functional and DFT-D3 dispersion corrections~\cite{Grimme2016}. We compensated for the underestimation of the DFT bandgap by applying a scissor operator to match the gap calculated from previous first-principles many-body perturbation theory within the GW approximation~\cite{Hedin1965} in the sHHG simulation calculations~\cite{ataei2021}. To model higher pressure conditions, the resulting relaxed lattice parameters $a$ and $c$ were uniformly compressed by relative amounts consistent with prior theoretical and experimental studies (see Methods)~\cite{Hromadova2013,Bandaru2014,Fan2015}.

\begin{figure}
    \centering
    \includegraphics[width=1\textwidth]{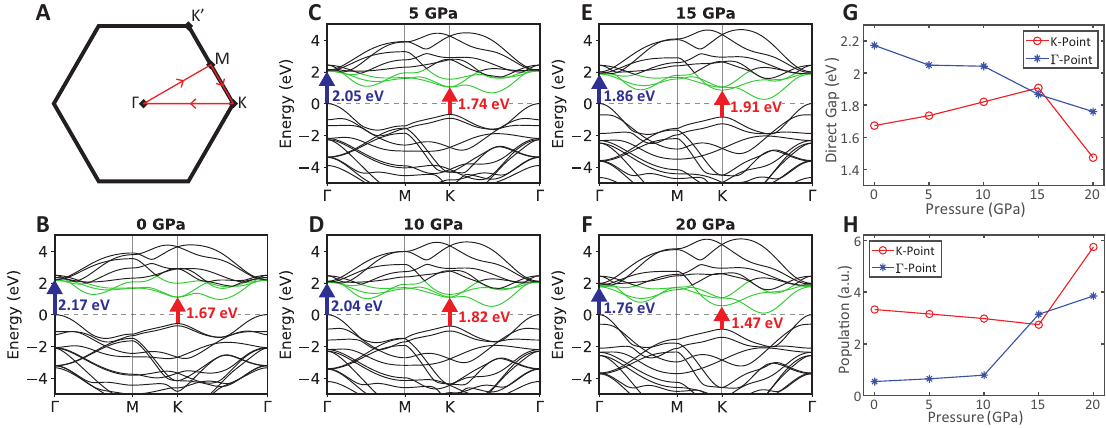}
    \caption{
    \textbf{MoS$_2$ band structure change with pressure.}
    The DFT band structure of MoS$_2$ along the path shown in (\textbf{A}) at (\textbf{B}) ambient pressure, (\textbf{C}) 5~GPa, (\textbf{D}) 10~GPa, (\textbf{E}) 15~GPa, and (\textbf{F}) 20~GPa calculated using the lattice parameters given in Table~S1.
    The direct bandgaps at \textbf{K} and $\mathbf{\Gamma}$ are denoted with red and blue vertical arrows respectively. The three lowest-energy CBs where band inversion occurs are highlighted in green.
    (\textbf{G}) shows the trend of the direct bandgaps at \textbf{K} (red) and $\mathbf{\Gamma}$ (blue) with pressure.
    (\textbf{H}) shows the trend in the time-integrated population of the first CB at the \textbf{K} (red) and $\mathbf{\Gamma}$ (blue) with pressure. The population was integrated over the first 200 fs during the simulated sHHG process (see Fig.~S2 for full time traces). }
    \label{fig:bandstructure}
\end{figure}

We observed a decrease in the indirect bandgap as a function of the applied pressure from nearly 1~eV at ambient pressure to about 0.1~eV at 20~GPa (Table~S1), consistent with previous calculations~\cite{Nayak2014}.
The calculated band structures of bulk MoS$_2$ at different pressures are shown in Fig.~\ref{fig:bandstructure}B-F, revealing significant pressure-induced modifications arising from lattice strain.
One of the clearest measures of these changes can be seen by comparing the direct bandgaps at the $\mathbf{\Gamma}$- and \textbf{K}-points (Fig.~\ref{fig:bandstructure}G), which start at 2.2~eV and 1.7~eV at ambient pressure respectively, at the DFT level without a scissor shift, and proceed to decrease and increase, respectively, with pressure. These changes can be understood from the orbital character of the band edge at different high-symmetry points (Fig.~S1). The split valence band at $\mathbf{\Gamma}$ consists of a pair of bonding and anti-bonding states arising from the interaction of sulfur $p_z$-orbitals in neighboring layers. As the pressure increases, the overlap of the neighboring $p_z$-orbitals increases, pushing up the valence band edge and decreasing both the direct gap at $\mathbf{\Gamma}$ and the overall indirect gap. Concomitantly, the direct bandgap at \textbf{K}, which consists primarily of states of Mo $d$-orbital character, increases with pressure, perhaps due to the increasing quantum confinement experienced by the electrons, which are largely confined to the Mo layer. 
At a pressure of 15~GPa, the order of the two direct gaps at $\mathbf{\Gamma}$ and \textbf{K} switch, with a value of 1.86~eV at $\mathbf{\Gamma}$ and 1.91~eV at \textbf{K}, resulting in a significant qualitative transition in the electronic structure.
This transition in k-space of the lowest direct bandgap from \textbf{K} to $\mathbf{\Gamma}$ has important implications for the sHHG mechanism, particularly in the context of the three-step model illustrated in Fig.~\ref{fig:DACschematic}C. The initial step---tunnel excitation---is generally understood to occur across a direct bandgap, with its efficiency highly sensitive to the magnitude of the gap~\cite{SingYou2017}.
Therefore, when the direct gaps at $\mathbf{\Gamma}$ and \textbf{K} cross in energy, tunnel excitation rates are predicted to become similar at both critical points, leading to a transition in the sHHG mechanism where many more carrier trajectories become equally accessible, potentially causing significant changes to emissions such as those observed experimentally around 15~GPa.

To fully understand how the pressure-induced transition in the band structure affects the field-driven nonlinear currents critical in sHHG, we performed a first principles simulation of the time evolution of the density matrix as described in the Methods, consistent with prior approaches~\cite{chan2021giantexciton, Kemper2013, Tancogne-Dejean2017a, Yue2022signatures, Yue2022intro, ChangLee2024}.
At ambient pressure, the smallest direct bandgap occurs at the \textbf{K}-point, leading to dominant carrier generation there as shown by the time-integrated population in Fig.~\ref{fig:bandstructure}H (full population dynamics in Fig.~S2). The resulting nonlinear sHHG current therefore primarily originates from the region in k-space around the \textbf{K}-point, as illustrated by the simulated current shown in Fig.~S3.
As pressure increases and the direct gaps at $\mathbf{\Gamma}$ and \textbf{K} decrease and increase respectively, the relative carrier population at $\mathbf{\Gamma}$ grows along with its relative contribution to the time-dependent sHHG current. By 15~GPa, carrier populations at \textbf{K} and $\mathbf{\Gamma}$ become similar due to their degenerate direct gaps, leading to the contributions to sHHG current from these two critical points to also become similar. However, rather than adding to each other, we show that the nonlinear current in these two regions is out of phase (Fig.~S3D inset), leading to deconstructive interference~\cite{Cao2021} that decreases the total sHHG current (Fig.~S3F).
The resulting decrease in sHHG intensity and modification of the anisotropy—seen clearly at 14.5~GPa in the experimental data—coincides with the calculated crossover of the smallest direct gap from \textbf{K} to $\mathbf{\Gamma}$ and the resulting destructive interference in the sHHG current, indicating a strong link between the evolving band structure and emission behavior through this electronic transition.

When the pressure in the simulations is further increased to 20~GPa, the lowest conduction bands highlighted in green in Fig.~\ref{fig:bandstructure} undergo reordering at \textbf{K}, with the third lowest CB at ambient pressure (CBM+2) becoming the lowest conduction band at 20~GPa. As shown by the PDOS (Fig.~S1), this reordering is caused primarily by a decrease in energy of CBM+2, which has strongly hybridized S $p$-orbital and Mo $d$orbital character, corresponding to the Mo-S bond. This leads to the \textbf{K}-point having a significantly smaller direct bandgap at 20~GPa, leading to a return to an sHHG mechanism in which carriers tunnel-excited at the \textbf{K}-point are dominant. This is demonstrated by the relative increase in CB population at \textbf{K} at 20~GPa (Fig.~\ref{fig:bandstructure}H) in the simulations. The restructuring of the lowest CBs around the \textbf{K}-point leading up to 20~GPa and associated changes in carrier dynamics in the sHHG mechanism correlate with the additional changes in the sHHG emissions around 20~GPa including a second 30$^\circ$ rotation in the 9$^{\text{th}}$ and 11$^{\text{th}}$ harmonics (Fig.~\ref{fig:pressureAnisotropy}).
The shift back to a single dominant direct bandgap minimum at \textbf{K} also reduces the number of energetically degenerate carrier trajectories, mitigating destructive interference in the time-dependent sHHG current at 20~GPa (Fig.~S3) and explains the observed recovery of sHHG intensity and clearer six-fold anisotropy at higher pressures. In addition, the continued decrease of the smallest direct bandgap should continue to increase overall sHHG efficiency in the absence of significant restructuring of the band structure and agrees with the trend of increasing sHHG signal up to 30~GPa.

\subsection*{Theoretical reproduction of experimental sHHG anisotropy}

\begin{figure}
    \centering
    \includegraphics[width=1\textwidth]{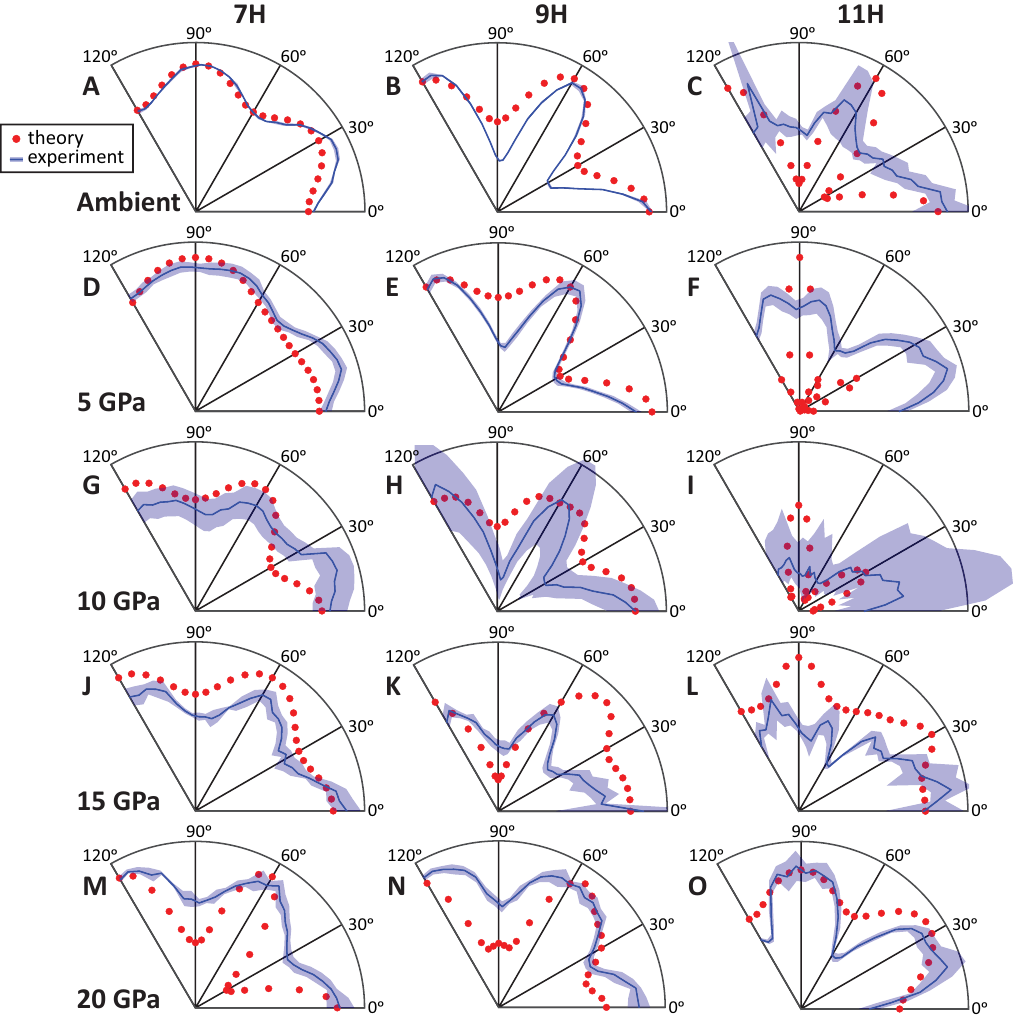}
    \caption{
    \textbf{Simulated sHHG anisotropy compared to experimental sHHG anisotropy.}
    At ambient pressure conditions, the anisotropy of the 7$^{\text{th}}$ (\textbf{A}), 9$^{\text{th}}$ (\textbf{B}), and 11$^{\text{th}}$ (\textbf{C}) harmonics are shown, with theoretical data as red dots and experimental data as a blue line with the shaded region representing a standard deviation of several measurements.
    (\textbf{D})-(\textbf{F}) show the 7$^{\text{th}}$, 9$^{\text{th}}$, and 11$^{\text{th}}$ harmonics respectively at 5~GPa.
    (\textbf{G})-(\textbf{I}) show the same harmonics at 10~GPa, (\textbf{J})-(\textbf{L}) show the same harmonics at 15~GPa, and
    (\textbf{M})-(\textbf{O}) show the same harmonics at 20~GPa.
    }
    \label{fig:compareThoeryExp}
\end{figure}

To further confirm that the observed changes in sHHG are indeed driven by the pressure-induced electronic structure evolution and associated population and current dynamics, we compare the modeled harmonic emissions directly with experimental data (Fig.~\ref{fig:compareThoeryExp}). The simulations reproduce key experimental features, most notably the shift in anisotropy of the 7$^{\text{th}}$ harmonic, which exhibits a breakdown of six-fold symmetry at a few~GPa and a 30$^\circ$ rotation of its emission maxima by 15~GPa—from 30$^\circ$ and 90$^\circ$ to 0$^\circ$, 60$^\circ$, and 120$^\circ$. Similar but more dramatic symmetry loss is observed in the 9$^{\text{th}}$ and 11$^{\text{th}}$ harmonics, consistent with the greater number of energetically-degenerate carrier trajectories available at higher harmonic orders~\cite{Yue2022signatures}. The simulation captures both the pronounced transition in anisotropy leading up to 15~GPa and the more gradual modifications at higher pressures. These effects are directly attributable to the evolving population dynamics that govern the sHHG current, and together, the strong agreement between simulation and experiment underscores the sensitivity of sHHG to subtle changes in electronic structure in the absence of a structural phase transition.

In summary, the evolution of sHHG in MoS$_2$ under pressure can be understood through the interplay of band structure changes and field-induced population dynamics. Near ambient pressure, sHHG is dominated by tunnel excitation at the \textbf{K}-point, producing the characteristic six-fold anisotropy. As pressure increases, the direct bandgap at the $\mathbf{\Gamma}$-point narrows, leading to comparable carrier populations at both \textbf{K} and $\mathbf{\Gamma}$ by 15~GPa. This degeneracy introduces multiple excitation pathways, enhancing deconstructive interference in the nonlinear current, which suppresses the sHHG signal and distorts the anisotropy. The resulting minimum in emission intensity and 30$^\circ$ anisotropy rotation near 15~GPa are signatures of this electronic crossover. At higher pressures, further warping of the conduction bands restores \textbf{K}-point dominance, reducing interference and leading to a recovery of sHHG intensity and a return toward six-fold symmetry. By 30~GPa, the increased harmonic yield is consistent with a reduced bandgap and the onset of semimetallic character, highlighting the sensitivity of sHHG to subtle electronic transitions in the absence of structural phase change.
Each of these subtle pressure-induced changes to the band structure of MoS$_2$ up to 30~GPa have not been accessible by other experimental techniques to date~\cite{Aksoy2006, Hromadova2013, Bandaru2014, Fan2015}.
Our use of sHHG has enabled the observation of an isostructural electronic transition in MoS$_2$ that had not been previously detected.

In conclusion, our findings establish that sHHG in DACs can overcome a long-standing barrier – the inability to probe electronic structures under elevated pressures. By capturing band dispersions and electronic transitions optically at extreme pressures in a DAC, sHHG provides unprecedented access to quantum phases that were previously beyond reach. This advance cements sHHG as a transformative spectroscopic tool for high-pressure research, one that opens new frontiers in fundamental science and enables the exploration of pressure-tuned topological states, emergent superconductivity, and other exotic electronic phases with broad implications for quantum materials and future technologies.

\subsection*{Materials and Methods}

\subsubsection*{Sample preparation}

Diamond anvil high-pressure cells of custom design were equipped with diamonds in the Boehler-Almax geometry, having 250~$\mu$m culets.
DAC gaskets were formed by preindenting Re foil to a thickness 25--30~$\mu$m, followed by laser micromachining a sample chamber typically 180~$\mu$m in diameter.
Molybdenum disulfide (2H-MoS$_2$; synthetic, 2Dsemiconductors USA) samples were prepared \textit{via} mechanical exfoliation from a monocrystalline bulk~\cite{Novoselov2005}.
Bulk material was thinned through repeated application of adhesive tape, then applied to a gel substrate (Gel-Pak) to transfer a thin many-layer sample.
Suitable samples were identified based on contrast in an optical microscope, with the selection criterion being uniform thickness over the focal size of the MIR driver, \textit{i.e.} several 10s of $\mu$m.
All samples used were well within the bulk limit.
Selected flakes were transferred to the culet of a diamond \textit{via} a transfer technique using common nail polish.~\cite{Haley2021}
The diamond was cleaned in baths of acetone and isopropanol to remove nail polish residue.
He was inserted as a pressure transmitting medium using a high-pressure gas loading apparatus (Top Industrie)~\cite{Smith2018_gasload} and samples were sealed at 3~kbar.

\subsubsection*{Optical microscopy and in situ pressure measurement within a DAC}

\begin{figure}
	\centering
	\includegraphics[width=1\textwidth]{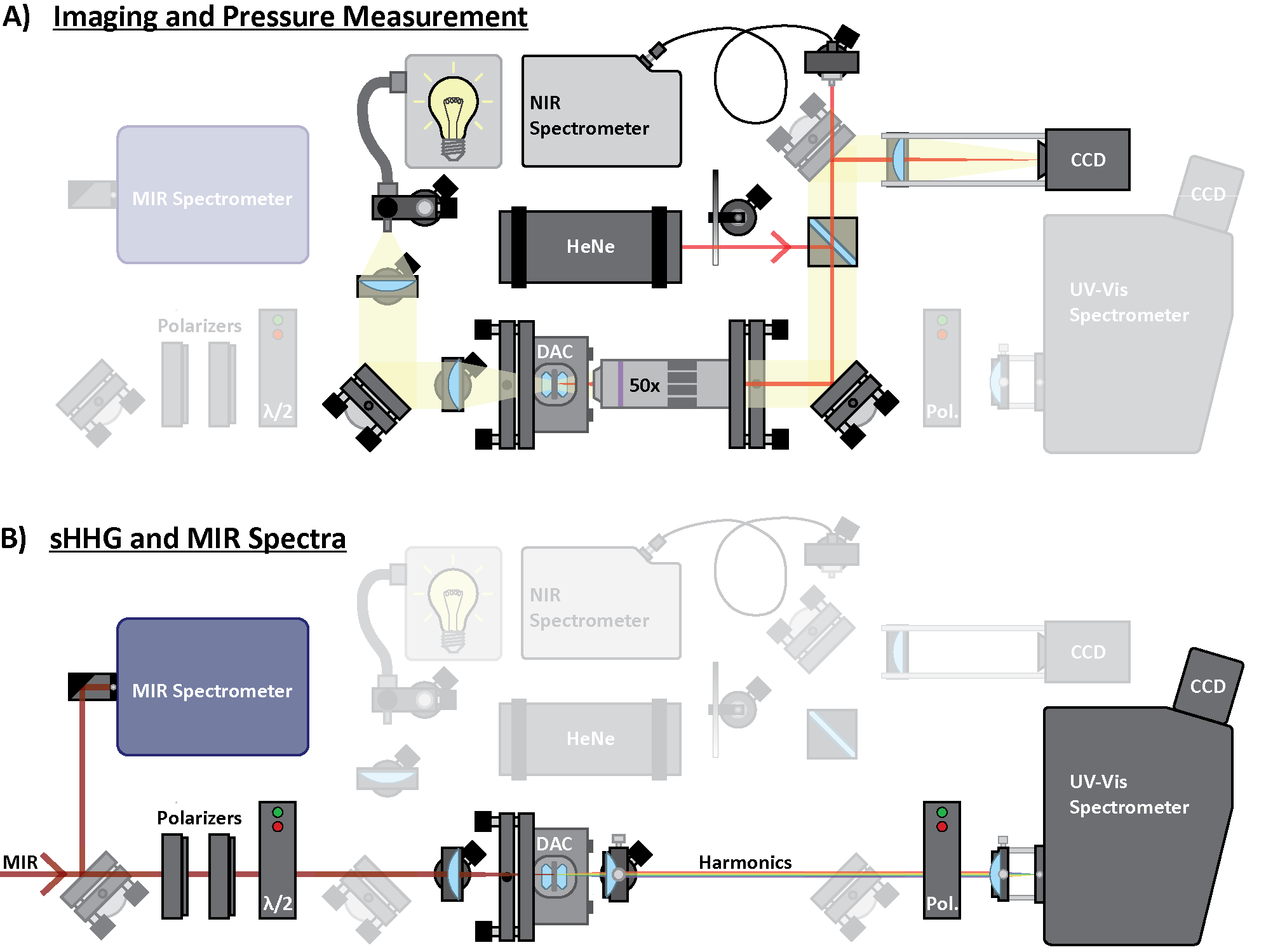}
	\caption{\textbf{sHHG spectroscopy setup.}
	(\textbf{A}) the setup used to take microscope images inside the DAC as well as measure ruby fluorescence for pressure calibration.
    Mirrors, the 50x objective, and the lens after the sample are on magnetic gimbal flip mounts to allow switching from the imaging configuration to the spectroscopy configuration (\textbf{B}).
    The spectroscopy setup was used to measure the MIR spectrum and the UV-vis sHHG spectra.}
	\label{fig:setup}
\end{figure}

For optical microscopy and in situ pressure measurements inside the DAC, a bright-field microscope was co-integrated into the sHHG setup (Fig.~\ref{fig:setup}A).
A white light source (Dolan-Jenner Industries, Fiber Lite 170-D) illuminates the DAC sample chamber in transmission geometry. A 50x objective (OptoSigma, PAL-50-NUV-A) combined with a CCD camera (ImagingSource, DMK 33G445) enables optical microscopy of the sample region.
This setup is used for in situ pressure measurements inside the DAC using ruby fluorescence spectra. For the pressure measurement, a glass beam splitter reflects $\sim$5\% of a helium-neon (HeNe) laser centered at 633~nm through the 50x objective, focusing the beam on a single ruby in the DAC sample chamber. Fluorescence of the ruby was collected in the objective and $\sim$95\% of the signal passed through the beam splitter before it was fiber-coupled to a high-resolution NIR spectrometer (OceanOptics, HR-6VN750-10). The ruby fluorescence is significantly red-shifted from the HeNe output (690-710~nm compared to 633~nm), allowing the spectrometer to measure the fluorescence spectrum with minimal interference from the HeNe. The pressure was determined by comparing the ruby fluorescence in ambient conditions on a sapphire substrate with the ruby fluorescence inside the DAC.
A standard calibration curve was used to determine the pressure inside the DAC~\cite{Dewaele2008}.

\begin{figure}
	\centering
	\includegraphics[width=0.5\textwidth]{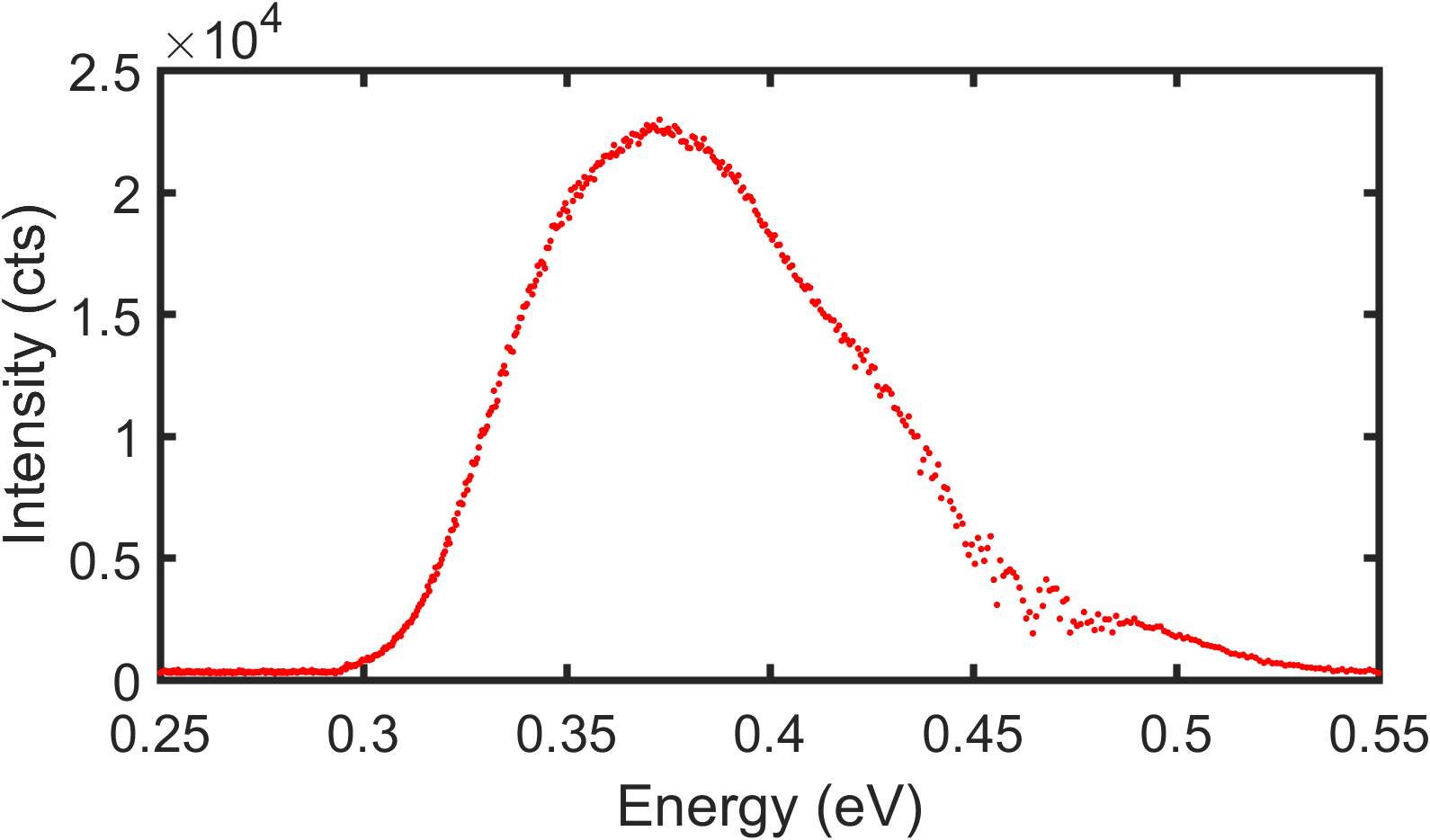}
	\caption{\textbf{MIR spectrum.}
	This spectrum shows the MIR pump used to generate harmonics from MoS$_2$ inside a DAC.
    The center of the MIR spectrum is at 0.37$\pm$0.01~eV and the FWHM is 0.10$\pm$0.01~eV. The pulse duration is approximately 100~fs previously determined using second-harmonic frequency-resolved optical gating (SHG-FROG) measurements.}
	\label{fig:MIRspect}
\end{figure}

\subsubsection*{sHHG spectroscopy and anisotropy instrument}

sHHG measurements were performed using a 100~fs mid-infrared (MIR) pulse with a pulse energy of 500$\pm$30~nJ. The maximum pulse energy of the MIR driver is around 50-60~$\mu$J; this is attenuated by a pair of wire grid polarizers (Thorlabs, WP25H-C) for this work so as to not damage the MoS$_2$ flake. The MIR pulse was generated from a Legend Elite Duo Ti:sapphire amplifier (Coherent) operating at 1~kHz, producing 30~fs pulses at 800~nm. Approximately 4.5~W of the 14~W output was diverted and the pointing was stabilized using an active beam stabilization system (MRC Systems, 1kHz-CW) before being directed into a optical parametric amplifier (TOPAS Prime Plus, Light Conversion). The resulting signal and idler were combined via noncollinear difference frequency generation (NDFG Unit, Light Conversion) to produce the tunable MIR pulse used in this study.
The spectrum of this MIR pulse is shown in Fig.~\ref{fig:MIRspect}.
Despite the measured center energy of the MIR being 0.37$\pm$0.01~eV, harmonics seem to be selectively generated by the higher energy part of the spectrum, resulting in experimental harmonics that would suggest a center MIR energy of 0.38~eV.
Given that the harmonics are significantly narrower in energy and thus give a more precise determination of the effective photon energy of the driver, we considered the center MIR photon energy to be 0.38~eV for the theoretical calculations.

The DAC is inserted into the instrument on a three-axis stage with additional pitch-tilt alignment to ensure the compression axis of the DAC is coaxial with the optical beam path.
The outer diamond surface of the DAC was aligned to be perpendicular to the MIR beam path using the back reflection of a HeNe alignment laser (Thorlabs, HNL020L) aligned along the same path so that the MIR field only exhibits in-plane components at the sample.
Additionally, the DAC was oriented such that the MIR beam passed first through the diamond/MoS$_2$ interface and then through the MoS$_2$/He interface so that the harmonics would be generated and measured primarily at the rear MoS$_2$/He interface (Fig.~\ref{fig:DACschematic}B).
This approach was chosen because diamond, as the substrate, is more likely to interact with MoS$_2$ and potentially modify interfacial properties while the interface with He is expected to represent unaltered MoS$_2$.

For sHHG measurements in the DAC (detailed setup in Fig.~\ref{fig:setup}B),  the MIR pulses were focused using a MIR anti-reflection coated CaF$_2$ lens (Thorlabs, LA5315-E1) such that the spot size on the sample was 30~$\mu$m, giving a field strength of 0.71$\pm$0.08~TW/cm$^2$ for the measurements of MoS$_2$.
Signal was collected in a CaF$_2$ lens (Thorlabs, LA5763) and spectra were measured using a spectrometer (Andor, Kymera 328i, gratings SR-GRT-300-300 and SR-GRT-150-500) coupled to a CCD camera that is kept at low temperature of -75$^\circ$C (Andor, iDus DU420A-BU2).
For anisotropy measurements, a motorized rotation stage (Thorlabs, K10CR1) manipulates a MIR $\lambda$/2 waveplate (B. Halle, $\lambda$/2 achromatic 3-6~$\mu$m) to adjust the polarization before the sample. A co-rotating wire grid polarizer (Thorlabs, WP25M-VIS) behind the sample allows for selecting sHHG emissions with either parallel or perpendicular polarization w.r.t. the incident MIR driving field.

Often, the parallel and perpendicular components of sHHG anisotropy are analyzed separately because some additional information can be extracted by analyzing the typically much weaker perpendicular component of sHHG emissions~\cite{Liu2017, Yue2022signatures, ChangLee2024}.
However, for the analysis here, we simply add the parallel and perpendicular emissions together to attain the total sHHG at each driver angle relative to the sample for the anisotropy data. This is equivalent to removing the second polarizer from the setup and measuring the total emission yield.
This simplifies our analysis by avoiding complications from any birefringence of the diamond in the DAC. It is known that diamond can exhibit weak anomalous birefringence~\cite{Eaton-Magana2016}, which could lead to a slight rotation and small elliptical component to the MIR driver after it passes through the DAC. It has been shown experimentally and theoretically that sHHG emissions from MoS$_2$ go to zero with increasing ellipticity~\cite{Lou2020}, thus a small elliptical component of the MIR driver can be considered to not contribute to the signal. A rotation of the linear polarization of the MIR driver would lead to contamination between parallel and perpendicular components of the sHHG emissions because the effective polarization of the MIR driver would not be perfectly parallel or perpendicular to the second polarizer. Therefore, we did not analyze the parallel and perpendicular components separately, but rather analyzed the total sHHG emission at each driver angle relative to the crystal structure of MoS$_2$ to avoid any issues accounting for birefringence. In fact, this also eases the comparison with theoretical calculations as the parallel and perpendicular components of the calculated current do not have to be separated.
To define our zero-angle relative to the crystal structure independent of any rotation from the diamond, we measured sHHG anisotropy from the MoS$_2$ flake at ambient pressure inside the DAC. The anisotropy agreed well with previous measurements of few-layer and monolayer MoS$_2$~\cite{Liu2017, Lou2020, Yue2022signatures}, which we used to assign the zero-angle relative to the crystal structure as defined in Fig.~\ref{fig:DACexample}D.

\subsubsection*{sHHG of MoS$_2$ inside a DAC}

\begin{figure}
	\centering
	\includegraphics[width=1\textwidth]{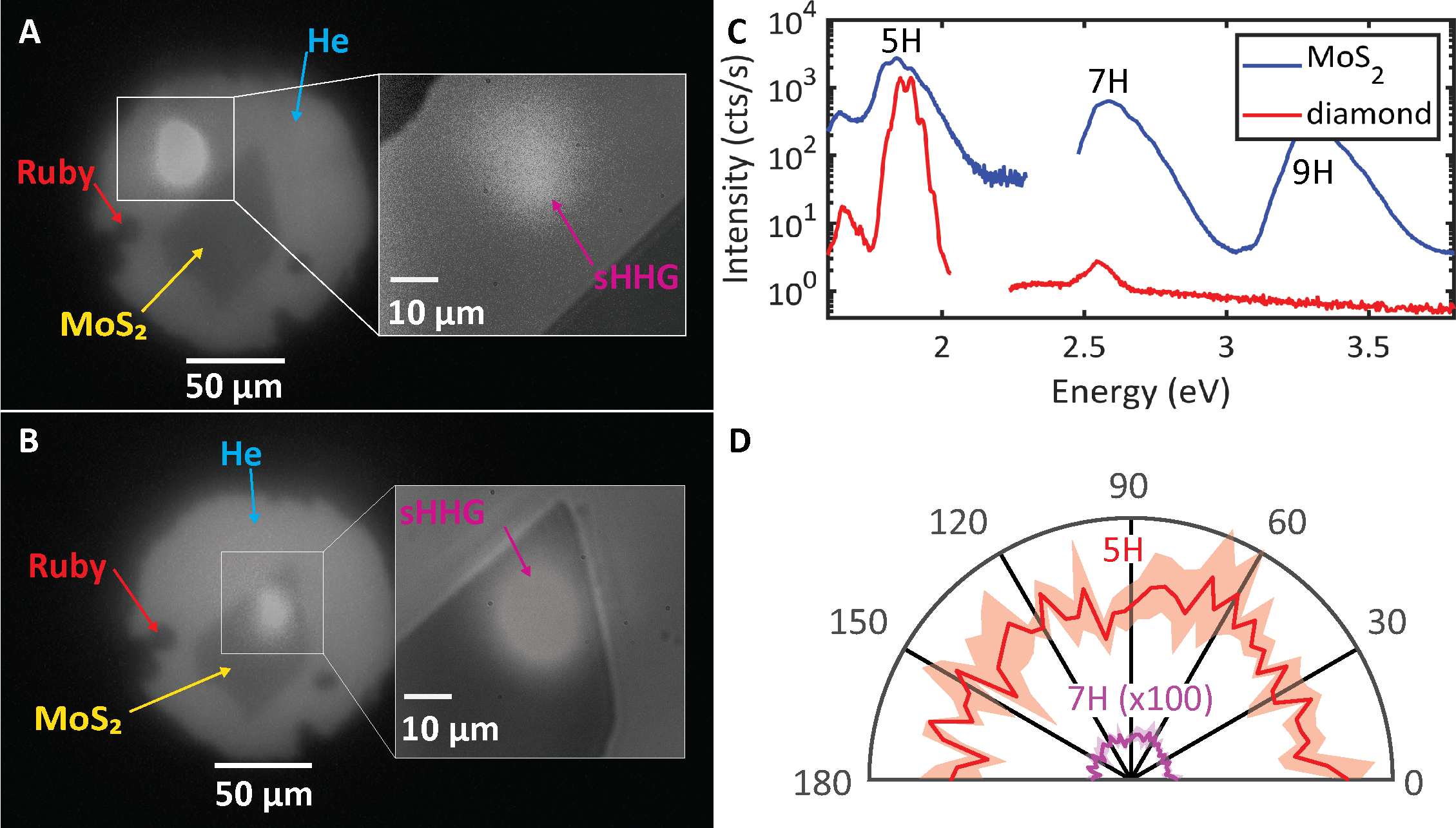}
	\caption{\textbf{Background sHHG from diamond and solid He at 30 GPa.}
    Optical microscope image through the DAC with the MIR driver incident on (\textbf{A}) an empty area of the DAC and (\textbf{B}) the MoS$_2$ sample. The insets are taken using a 50x objective and used the same integration time of the CCD camera of 0.5~s.
    (\textbf{C}) sHHG spectrum of the empty area of the DAC (red) and MoS$_2$ sample inside the DAC (blue) at 30~GPa.
    The MIR pump used was the same as used for MoS$_2$ throughout the manuscript, with an intensity of 0.71$\pm$0.08~TW/cm$^2$ for both spectra.
    (\textbf{D}) anisotropy of the 5$^{\text{th}}$ (red) and 7$^{\text{th}}$ (purple) harmonics from the empty area of the DAC. The shaded region represents a standard deviation of several measurements.}
	\label{fig:backgroundSignal}
\end{figure}

A critical aspect of this work is to confirm that observed sHHG emissions stem from the sample material and not parts of the DAC that are necessarily traversed by the MIR beam. To confirm this, measurements with the MIR beam only passing through diamond and He in the DAC were taken, \textit{i.e.} at a location in the sample chamber away from the MoS$_{2}$ sample. Fig.~\ref{fig:backgroundSignal}A and B show the optical microscope image of sHHG emissions from the empty area of the DAC compared to the MoS$_2$ sample. The inset images are taken with a 50x objective and have the same integration time, showing that the emissions from the empty area of the DAC are much weaker despite the same intensity of the MIR driver of 0.71$\pm$0.08~TW/cm$^2$.
The emissions from the empty area of the DAC were characterized and some 5$^{\text{th}}$ and very weak 7$^{\text{th}}$ harmonics were observed (Fig.~\ref{fig:backgroundSignal}C). These measurements were performed at all pressures including those where He becomes solid. The observed backgrounds were constant across used pressure ranges indicating that the background emissions likely stem from diamond. We note that similar sHHG spectra were observed from polycrystalline diamond in a previous study of large bandgap materials carried out in the same setup, though in the DAC measurements here a much lower intensity MIR driver is used, leading to less total emissions from diamond~\cite{Korican-Barlay2024}.
Both of the harmonics observed from diamond are isotropic under the MIR driver conditions used to study MoS$_2$ (Fig.~\ref{fig:backgroundSignal}D), meaning that they should not affect the analysis of the anisotropy from MoS$_2$.
However, the 5$^{\text{th}}$ harmonic from diamond is comparable in intensity to the 5$^{\text{th}}$ harmonic from MoS$_2$, therefore, to avoid potential complications we chose to exclude this harmonic from our analysis.
On the other hand, the 7$^{\text{th}}$ harmonic from diamond was found to be three orders of magnitude weaker than that from MoS$_2$ (Fig.~\ref{fig:backgroundSignal}C). With its weak intensity relative to signal from the sample and its isotropic behavior, the 7$^{\text{th}}$ harmonic from diamond was negligible in the analysis of MoS$_2$. No higher harmonic orders were observed from diamond under the MIR driver condition used in this work (Fig.~\ref{fig:backgroundSignal}C) which agrees with our previous study of diamond~\cite{Korican-Barlay2024}, meaning that sHHG from MoS$_2$ of 9$^{\text{th}}$ order and greater are free of any background contribution.

In addition, to confirm the typical non-perturbative mechanism for sHHG described in Fig.~\ref{fig:DACschematic}C for the sample studied here, the power dependencies of the harmonics from MoS$_2$ at 30~GPa were measured. The results show the 5$^{\text{th}}$, 7$^{\text{th}}$, and 9$^{\text{th}}$ harmonics follow power laws of orders between 3-4 (Fig.~S4). The power law orders of the harmonics are similar, independent of harmonic order, and all less than expected from perturbation theory, which would give power law orders of 5, 7, and 9 for the 5$^{\text{th}}$, 7$^{\text{th}}$, and 9$^{\text{th}}$ harmonics, respectively. This demonstrates that the observed sHHG from MoS$_2$ inside the DAC is indeed non-perturbative consistent with previous works in the field~\cite{Ghimire2011, Liu2017}, validating our analysis of the sHHG emissions from MoS$_2$ at extreme pressure.

\subsubsection*{Electronic structure and sHHG simulations}

Density functional theory (DFT) calculations were performed using the \textsc{Quantum Espresso} software package~\cite{Giannozzi_2009}.
The Perdew, Burke, and Ernzerhoff (PBE) generalized grandient approximation (GGA) functional was used to approximate the exchange and correlation effects~\cite{Perdew1997}.
Norm-conserving pseudopotentials obtained from the ONCV SG15 library were used to replace the core electrons~\cite{Scherpelz2016}. Dispersion corrections were included in the simulation using the Grimme DFT-D3 method~\cite{Grimme2016}.
We found that a plane wave cutoff energy of 80~Ry and a 12\texttimes12\texttimes4 k-grid converged the total energy of our DFT calculation within 0.1~meV.

DFT simulations of high-pressure structures were performed by reducing the fully relaxed DFT lattice parameter by the corresponding percentage obtained from previous experimental and theoretical studies~\cite{Hromadova2013,Nayak2014,Fan2015}, followed by an additional relaxation of the atomic positions after the lattice constraint.

sHHG simulations were performed by solving the equation of motion for the electron density matrix as shown in Equation \ref{eq:time_density} where $H_0$ is the mean field DFT Hamiltonian at equilibrium corrected with a scissor operator to account for the DFT underestimation of the band gap; $\mathbf{E}(t)$ is the external electric field at time $t$, which couples to the system in the length gauge through the position operator $\mathbf r$. $n$ and $m$ are the band indices and $\textbf{k}$ is a $\textbf{k}$-point in the Brillouin zone. We used a $36 \times 36 \times 9$ $\textbf{k}$-point grid including 14 valence bands and 8 conduction bands in our time evolution calculations. The total time of the time evolution was 200~fs with a time step of 0.05~fs. Depopulation is described by a diagonal term of 25~meV obtained from previous work~\cite{Taghizadeh2019}. Dephasing is described by a off-diagonal term of 200~meV which is in the range of previous works~\cite{Taghizadeh2019,Attacalite2013}.
A locally smooth gauge as implemented in our previous work is used to treat the intraband coupling terms, $[r^{intra},\rho]_{nm\mathbf{k'}}$~\cite{chan2021giantexciton}.

\begin{equation}
    i\hbar \frac{\partial}{\partial t} \rho_{nm,\textbf{k}} (t) = [H_0  -e\textbf{E} \cdot {\textbf{r}}, \rho]_{nm,\textbf{k}} 
    \label{eq:time_density}
\end{equation}

\noindent\ The time-dependent current is found by taking the trace of the time-dependent density matrix with the velocity operator $\textbf{v}$ as shown in Equation~\ref{eq:tdCurrent}. An example of a component of the time-dependent current responsible for the 9th harmonic is shown in Fig.~S3.

\begin{equation}
    \textbf{J} (t) = \textbf{Tr}(\rho(t)\textbf{v}),
    \label{eq:tdCurrent}
\end{equation}

\noindent\ The corresponding high harmonic contributions to the current are computed by taking the Fourier transform of the time-dependent current~\cite{Kemper2013,Tancogne-Dejean2017a,Yue2022intro,ChangLee2024} as shown in Equation \ref{eq:hhg}.

\begin{equation}
    \textrm{HHG}(\omega) = |\omega\int dt\, \textbf{J} (t) e^{-i\omega t}|^2
    \label{eq:hhg}
\end{equation}

\noindent\ The time-dependent population (as seen in Fig.~S2) information is obtained from the diagonal elements of the time-dependent density in Equation 1.
The population of the first CB at \textbf{K} (blue) and $\mathbf{\Gamma}$ (red) are shown in Fig.~S2 and are integrated to give the trend in Fig.~\ref{fig:bandstructure}F.


\clearpage

\bibliography{sHHG_DAC_MoS2_bib01}
\bibliographystyle{sciencemag}


\paragraph*{Funding:}
B.R.N. acknowledges the funding from the National Science Foundation Graduate Research Fellowship Program (DGE 1752814).
J.A.S. acknowledges the support from the Arnold O. Beckman Postdoctoral Fellowship Program.
B.R.N. and J.A.S. further acknowledge funding by the UC Office of the President within the Multicampus Research Programs and Initiatives (M23PR5931).
C.P.S. was supported by the U.S. Department of Energy, Office of Basic Energy Sciences through the condensed phase and molecular science program under contract No. DE-SC0023355 for clean energy manufacturing.
D.S. was supported by the National Science Foundation EPSCoR Research Infrastructure Improvement (RII) Track-4: EPSCoR Research Fellows under Grant No. 2327363.
M.W.Z. acknowledges funding from the Rose Hills Foundation, the Camille and Henry Dreyfus Foundation, the Hellman Foundation, and the W.M. Keck Foundation that enabled building different parts of the apparatus.
M.W.Z. further acknowledges funding from the National Science Foundation (NSF-DMR 2247363).

\paragraph*{Author contributions:}
B.R.N., V.C.L., D.S., and M.W.Z. wrote the original manuscript draft. Conceptualization of the research was carried out by B.R.N., J.A.S., C.P.S., D.S., and M.W.Z. The experimental investigation was performed by B.R.N., J.A.S., R.M.S., and D.S. The numerical simulations were carried out by V.C.L. and D.Y.Q. Methodology was developed by B.R.N., V.C.L., J.A.S., R.M.S., C.P.S., D.S., D.Y.Q., and M.W.Z. Funding was acquired by D.S., D.Y.Q., and M.W.Z. Visualization was done by B.R.N., V.C.L., and D.S. Project administration was led by M.W.Z. Supervision was provided by D.Y.Q. (theory) and M.W.Z. (experiment and overall). All authors discussed the results and contributed to final editing of the manuscript.

\paragraph*{Competing interests:}
There are no competing interests to declare.

\paragraph*{Data and materials availability:}
All data used for interpretation of the results is shown in the manuscript. Additional data and raw data can be requested from the corresponding author upon request.


\end{document}